# Influence of tangential displacement on the force of adhesion between a parabolic profile and plane surface

Valentin L. Popov [1,2,3,*], Iakov A. Lyashenko [1,4] and Alexander E. Filippov [1,5]

[1]Technische Universität Berlin, Institut für Mechanik, FG Systemdynamik und Reibungsphysik, Sekr. C8-4, Raum M 122, Straße des 17. Juni 135, 10623 Berlin, Germany
[2]National Research Tomsk State University, 634050 Tomsk, Russia
[3]National Research Tomsk Polytechnic University, 634050 Tomsk, Russia
[4]Sumy State University, 40007 Sumy, Ukraine
[5]Donetsk Institute for Physics and Engineering, National Academy of Science, 83114 Donetsk, Ukraine

*Corresponding author: v.popov@tu-berlin.de

**Abstract**

The force of adhesion of a rotationally symmetric indenter and an elastic half-space is analyzed analytically and numerically using an extension of the method of dimensionality reduction (MDR) for superimposed normal/tangential adhesive contacts. In particular, the dependency of the critical adhesion force on the simultaneously applied tangential force is obtained and the relevant dimensionless parameters of the problem are identified. The developed method is applicable straightforwardly to adhesive contacts of any bodies of revolution.

*Keywords:* adhesion, friction, tribology, shear force, numerical simulation, method of dimensionality reduction

## 1 Introduction

Johnson, Kendall and Roberts developed 1971 their classical theory of normal adhesive contact between two parabolic, isotropic elastic bodies ([1], JKR-theory) utilizing the analogy between the boundary of an adhesive contact and the tip of a crack in mode I (opening mode). They applied the same idea of the energy balance as Griffith used in his classical theory of cracks [2]. In the subsequent years, the theory of adhesive contacts developed rapidly [3] mostly utilizing various concepts developed in fracture mechanics.

In the JKR-theory – just as in the theory of Griffith – the equilibrium configuration of an adhesive contact is determined by minimizing the total energy of the system including the energy of elastic deformation of contacting bodies, the interface energy and the work of external forces [4]. As this energy does not depend on the tangential displacement, the adhesive contact formally has no "tangential strength". This apparently contradicts experimental observations. The contradiction is due to the microscopic heterogeneous structure of any interface (at the atomic scale at the latest) which provides final strength to a contact in the tangential direction either.

From the microscopic point of view, one can interpret the fracture condition of Griffith as requirement that the stress in a fixed distance (of atomic order) from the actual "crack tip" achieves some critical value ("stress criterion"). This condition leads to the macroscopic dependence of the critical stress as function of the crack length which is identical to relations obtained from the macroscopic energy balance [5]. Alternatively, one could require that the relative displacement of the faces of the crack achieve some critical value ("deformation criterion"). The stress and deformation criteria are equivalent for pure elastic bodies but can lead to different fracture conditions for elastomers [5]. In the present paper we will only deal with pure elastic bodies, so we can apply either the stress or the deformation criterion without loss of generality. Johnson studied 1997 the problem of adhesive contact under superimposed normal and tangential loading [6] and came to the conclusion that "when tangential forces are applied to an adhesive contact the consequences are not at all well understood". The situation has not change much till now.

In his paper of 1997 Johnson approaches the problem of adhesive contact under superimposed normal and tangential loading by considering the complete energy release rate at the boundary of an adhesive contact [6]

$$Q = \frac{1}{2E^*}\left[K_I^2 + K_{II}^2 + \frac{1}{1-\nu}K_{III}^2\right], \tag{1}$$

where $K_I$, $K_{II}$ and $K_{III}$ are the stress concentration factors defined as

$$K_{I,II,III} = \frac{F_{I,II,III}}{2a\sqrt{\pi a}} \tag{2}$$

with $a$ being the contact radius and $F_{I,II,III}$ the components of the applied force in the normal (*I*), and tangential directions (*II* – radial direction, and *III* – tangential to the boundary line direction). The problem of a circular contact remains axially symmetric only if Poisson number $\nu = 0$. In this case, the stress concentration factors for the modes *II* and *III* are equal along the entire boundary line. In the case of arbitrary Poisson number, Johnson suggest to evaluate the average values of $K_{II}$ and $K_{III}$ round the periphery of the contact area simplifying (1) to

$$Q = \frac{1}{2E^*}\left[K_I^2 + \frac{2-\nu}{2-2\nu}K_{II}^2\right]. \tag{3}$$

In terms of energy release rates, the condition of fracture can be formulated by equating the energy release rate to some critical value related with the work of adhesion $\Delta\gamma$. We would like to stress that this approach is by no means self-evident. Physically, it means that elastic energy parts due to normal and tangential loading contribute in equal manner to destroying the interfacial bonds. This may be true in some cases. For example, if the adhesion of surfaces is mediated by some polymer molecules which have to be destroyed by large enough elongation, then both normal and tangential deformations give contribution to the "fracture criterion". The same may be valid in a contact of atomically smooth surfaces with equal characteristic range of atomic interaction in normal and tangential direction. In this case, the tangential displacement of atoms at the interface due to "tangential part of elastic energy" will bring them already in a higher energetic position compared to those in the stress free state. Subsequently, a smaller work will is to be done by normal forces to complete the detachment. In this case, too, one can at least qualitatively assume that both normal and tangential parts of elastic energy give approximately the same contribution to the overall detachment energy. In other situations, however, this criterion can fail completely. Thus, if the characteristic range of atomic interactions in the in-plane direction is much smaller than in the normal direction, then the work of adhesion will be practically independent on the tangential loading and the criterion (3) will not be valid. One can also imagine a physical model in which there is some "microscopic friction" between the surfaces which are pressed to each other by relatively long-range von-der-Waals forces. In this case, the work of detachment will depend on the exact "direction of detachment". Thus, the true condition for the equilibrium of an adhesive contact under tangential loading cannot be determined from pure theoretical considerations as it may depend on detailed physics of interface strength.

Another important question in considering adhesion is what happens *after* the detachment took place at some position of the boundary of adhesive contact. If the detachment occurs due to *combined* action of normal and tangential loading, then it may well be the case that the adhesion bonds will be restored after the medium has relaxed the tangential part of elastic energy. This rebinding can actually take place or it can be prevented due to different reasons. The simplest one is a rapid change of the surface (for example, due to oxidation). Another reason may be due to irreversible changes of surface topography during detachment (so that the surfaces become incongruent and cannot restore the initial configuration. One further reason maybe that the actual work of detachment is much larger than the pure surface energy. In this case, the main part of elastic energy will disappear irreversibly and the relatively weak interface interactions will not be able to restore the integrity of the interface again. In all these cases we would have to do with an *irreversible adhesive contact*. In the present paper, we consider exactly this case of irreversible adhesion. One can interpret this case as a fracture problem of initially glued contact.

In the present paper we do not consider the interface physics in detail but just assume the validity of assumptions similar to those of Johnson, using the so-called method of dimensionality reduction (MDR). In a series of papers, Popov and co-authors have shown that contact problems of axially-symmetric tree-dimensional bodies can be equivalently represented by contacts with one-dimensional series of independent springs [5]. It is important to note, that the results for axially-symmetric contacts obtained with MDR are *exact* and not an approximation as often is believed. MDR was first proposed 2007 for non-adhesive contacts [7]. In his dissertation of 2011, Markus Heß found the MDR formulation also for adhesive contacts of bodies with arbitrary axis-symmetric form [8]. A short review of the MDR for contacts of bodies of revolution can be found in [9].

Let us briefly mention previous approaches to the problem of adhesion under superimposed normal and tangential load. In [10], the authors used discrete element method (DEM) for modeling contacts between cohesive, frictional particles with normal and tangential loading taking into account of adhesion forces between particles. In [11], a model of tangential adhesion contact was proposed, however under assumption that effects of normal and tangential force can be considered independently. In this investigation authors showed that in tangential contact problem the influence of adhesion can be approximately described in the terms of equivalent load. In set of articles of Guduru with co-authors [12], [13], [14], [15] the work of adhesion was considered as a function of "mode-mixing", which means that the work of adhesion depends on the direction of detachment [21], [22]. In [12], the theory of Guduru et. al. was verified experimentally. Tangential adhesion effects were investigated numerically within the framework of Coupled Eulerian-Lagrangian method in [16]. In [17], adhesion-induced plastic deformations due to tangential loading was considered. Tangential adhesion effects have been investigated in context of biological systems [18] and physics of particle interactions [10], [19], [20].

This paper is organized as follows. In Section 2 we recapitulate briefly the MDR approach for normal adhesive contacts and extend it for the case of superimposed normal and tangential loading under fixed-load and fixed-grips conditions. The model then is studied numerically and analytically and the dependence of the adhesive force on the tangential force is established in proper dimensionless variables. In Section 3 we describe numerical procedure in details and do comparison between numerical and analytical results. Section 4 concludes the paper.

## 2 MDR formulation for adhesive contact and analytical solution

### 2.1 MDR for normal adhesive contacts

In the case of rotationally symmetric bodies, the MDR consists of two simple steps [9]. First, the original three-dimensional profile $z = f(r)$ (shown in left subplot of the Fig. 1) is replaced by the one-dimensional profile $g(x)$ by means of the transformation:

$$g(x) = |x| \int_0^{|x|} \frac{f'(r)}{\sqrt{x^2 - r^2}} \, dr. \tag{4}$$

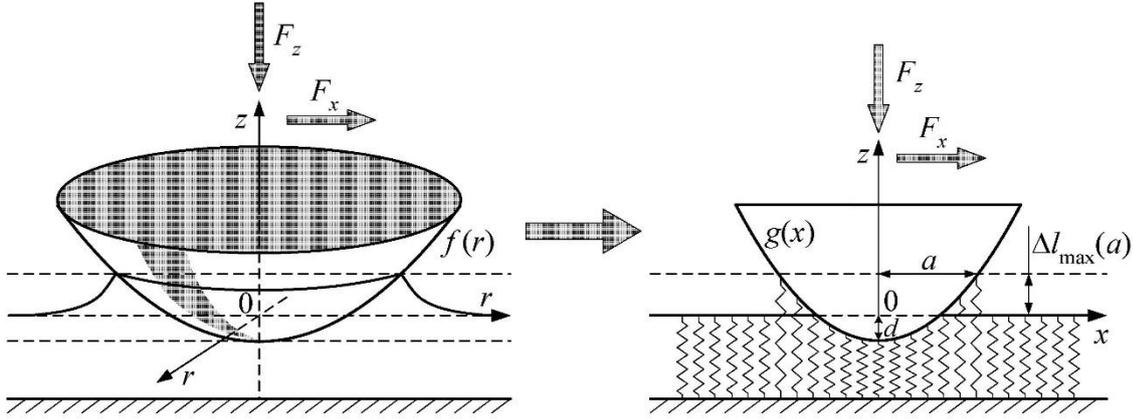

*Fig. 1. MDR transformation of the original three-dimensional profile $f(r)$ into one-dimensional image $g(x)$ and replacement of the elastic half-space by an elastic foundation. In presence of normal and tangential force and adhesion, the springs of the elastic foundation will be displaced both in normal and tangential direction. In this Figure, only vertical displacements are shown.*

If needed, the original surface $z = f(r)$ can be always restored from its MDR-transformed one-dimensional profile by

$$f(r) = \frac{2}{\pi} \int_0^r \frac{g(x)}{\sqrt{r^2 - x^2}} \, dx. \tag{5}$$

In the present paper, we will limit ourselves to parabolic profiles of the form $f(r) = r^2 / (2R)$. However, generalization to arbitrary other rotationally symmetric profiles is straightforward. In the case of parabolic profile, transformation (4) leads again to a parabolic profile $g(x)$ with a changed coefficient:

$$f(r) = \frac{r^2}{2R} \quad \Rightarrow \quad g(x) = \frac{x^2}{R}. \tag{6}$$

In the second step [9], the elastic half-space must be replaced by elastic foundation, as shown in Fig. 1, consisting of independent springs having normal and tangential stiffness

$$k_z = E^* \Delta x, \quad k_x = G^* \Delta x, \tag{7}$$

where $\Delta x$ is the spacing of the foundation and effective moduli $E^*$ and $G^*$ are defined as

$$E^* = \frac{E}{1-\nu^2} = \frac{2G}{1-\nu}, \quad G^* = \frac{4G}{2-\nu}, \tag{8}$$

so that

$$G^* = E^* \frac{2-2\nu}{2-\nu}. \tag{9}$$

Note that in the case of $\nu = 0$ both effective moduli coincide: $E^* = G^*$. For definiteness and simplicity, all numerical simulations and analytical calculations below are performed under this assumption.

We first recapitulate briefly the application of the MDR to normal adhesive contacts. If the MDR-transformed profile $g(x)$ is indented into the elastic foundation by the indentation depth $d$, then the displacement of individual springs inside the contact will be determined by the equation

$$u_z(x) = d - g(x) = d - \frac{x^2}{R}. \tag{10}$$

The size of the adhesive contact at the given indentation depth can be easily found from the principle of virtual work. The springs at the boundary of contact are stretched by $\Delta l = -u_z(a)$. The energy released through detachment of two boundary springs is equal to $E^* \Delta l^2 \Delta x$. Through detachment, the free surface area $2\pi a \Delta x \Delta \gamma$ is created (this energy can only be defined in the original, three-dimensional system). According to the principle of virtual work, the system will be in equilibrium if these two energies are equal:

$$E^* \Delta l^2 \Delta x = 2\pi a \Delta x \Delta \gamma. \tag{11}$$

It follows that the condition of equilibrium of boundary springs can be written as

$$\Delta l = \Delta l_{\max}(a) = \sqrt{\frac{2\pi a \Delta \gamma}{E^*}}. \tag{12}$$

This rule has been originally found by M. Heß in [8] and is known as *rule of Heß*. Combining (10) and (12) we get

$$u_z(a) = d - \frac{a^2}{R} = -\Delta l_{\max}(a) = -\sqrt{\frac{2\pi a \Delta \gamma}{E^*}}, \tag{13}$$

or

$$d = \frac{a^2}{R} - \sqrt{\frac{2\pi a \Delta \gamma}{E^*}}. \tag{14}$$

The normal force can be calculated as a sum of all spring forces:

$$F_z(a) = \int_{-a}^{a} u_z(x) \mathrm{d}x = 2E^* \int_0^a \left( d - \frac{x^2}{R} \right) \mathrm{d}x = \frac{4E^* a^3}{3R} - \sqrt{8\pi a^3 E^* \Delta \gamma}. \tag{15}$$

Later we will consider a more general situation where the indenter is displaced also in the tangential direction by $u_x^{(0)}$. It is convenient to present both analytical and numerical results in dimensionless units:

$$\tilde{a} = \frac{a}{a_0}, \quad \tilde{F}_z = \frac{F_z}{F_0}, \quad \tilde{d} = \frac{d}{d_0}, \quad \tilde{u}_x^{(0)} = \frac{u_x^{(0)}}{d_0}, \quad \tilde{u}_z = \frac{u_z}{d_0}, \quad (16)$$

where $F_0$, $a_0$ and $d_0$ are the critical values of the force, the contact radius and the absolute value of the indentation depth at the moment of detachment of the parabolic profile from the elastic half-space under "controlled force" condition [4]:

$$F_0 = \frac{3}{2}\pi R \Delta\gamma, \quad a_0 = \left(\frac{9\pi R^2 \Delta\gamma}{8E^*}\right)^{1/3}, \quad d_0 = \left(\frac{3\pi^2 R \Delta\gamma^2}{64 E^{*2}}\right)^{1/3}. \quad (17)$$

In dimensionless variables, equations (14) and (15) take the form

$$\tilde{d} = 3\tilde{a}^2 - 4\tilde{a}^{1/2} \quad (18)$$

and

$$\tilde{F} = \tilde{a}^3 - 2\tilde{a}^{3/2}. \quad (19)$$

These results coincide of course with the classical solution of Johnson, Kendall and Roberts [1]. The dependence of the dimensionless normal force on the dimensionless approach (indentation depth) implicitly defined by equations (18) and (19) is shown in Fig. 2 and will be used for testing numerical procedures described in Section 3.

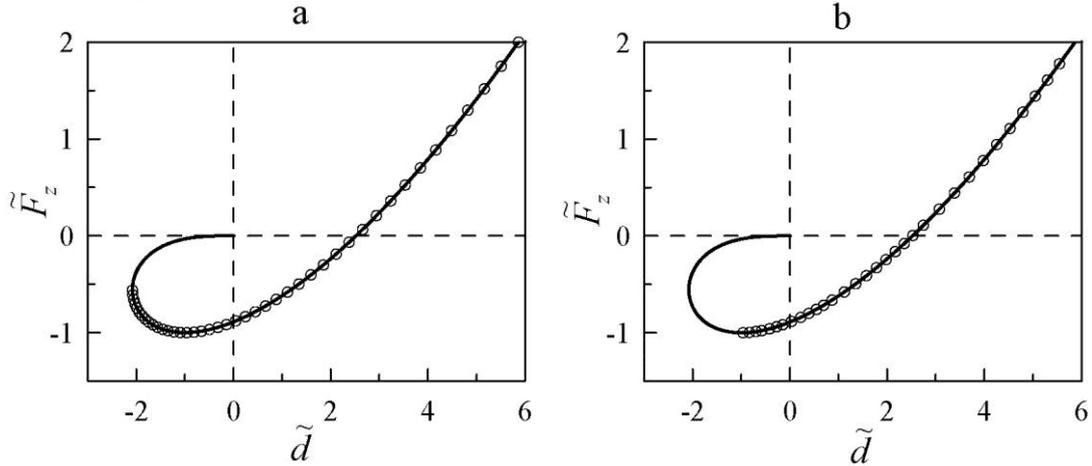

*Fig. 2: Dependence on the normal force on the indentation depth for the normal contact with adhesion. Solid lines describe analytical solution defined by equations (18) and (19). Circles present results of numerical experiments for "fixed-grips" (a) and "fixed-load" (b) conditions as described in Section 3.*

We now analyze the condition of instability of the contact, that means the conditions under which the possibility of the adhesive contact to sustain equilibrium gets lost. In doing this, we will consider two types of controlling the external loading conditions: "fixed-grips" and "fixed-load". Under "fixed-grips" condition is meant the control of the macroscopic displacement of the indenter via a very stiff external system. Physically this means that during the movement of the system toward the equilibrium state, the displacement is kept exactly constant. "Fixed load" conditions physically are realized by loading with the given force over very soft spring. Thus, in conditions of "fixed load", the force during the relaxation to the equilibrium remains fixed. The method described in this section has been used successfully for modeling influence of adhesion on impact between elastic particles, when "fixed-grips" condition was realized [20].

## 2.2 Superimposed normal and tangential loading

Let us now assume that the loading of the profile consists of superimposed normal force and tangential displacement $u_x^{(0)}$. The energy released by detachment of two boundary springs will now be equal to $E^* u_z(a)^2 \Delta x + G^* u_x^{(0)2} \Delta x$. Equating it to the work of adhesion $2\pi a \Delta x \Delta \gamma$, we will come to the following equilibrium condition

$$E^* u_z(a)^2 + G^* u_x^{(0)2} = 2\pi a \Delta \gamma. \qquad (20)$$

This rule is exactly equivalent to the rule obtained by Johnson on the basis of the energy release rate (3). From (20), for the elongation of the boundary springs we get

$$\left| u_z(a) \right| = \sqrt{\frac{2\pi a \Delta \gamma}{E^*} - \frac{G^*}{E^*} u_x^{(0)2}} \,. \qquad (21)$$

Using (10), we can write the relation between the indentation depth and contact radius in the form

$$d = \frac{a^2}{R} - \sqrt{\frac{2\pi a \Delta \gamma}{E^*} - \frac{G^*}{E^*} u_x^{(0)2}} \,. \qquad (22)$$

The normal and tangential forces are functions of contact radius $a$:

$$F_z = 2E^* \left( ad - \frac{a^3}{3R} \right), \qquad (23)$$

$$F_x = 2G^* a \cdot u_x^{(0)}. \qquad (24)$$

In dimensionless variables (16), Eqs. (22)-(24) can be written as

$$\tilde{d} = 3\tilde{a}^2 - \sqrt{16\tilde{a} - \frac{G^*}{E^*} \tilde{u}_x^{(0)2}} \,, \qquad (25)$$

$$\tilde{F}_z = \frac{\tilde{a}}{2} \left( \tilde{d} - \tilde{a}^2 \right), \qquad (26)$$

$$\tilde{F}_x = \frac{G^*}{2E^*} \tilde{a} \cdot \tilde{u}_x^{(0)} \,. \qquad (27)$$

These equations determine the normal force-indentation relation in the presence of tangential displacement. Note that substitution of (25) into (26) at $\tilde{u}_x^{(0)} = 0$ recapitulates the result (19) for the normal contact.

Let us stress that equation (20) assumes that the work of adhesion does not depend on the direction in what the surfaces are detached from each other. This is a physical assumption which may be incorrect in some systems [21], [22]. At this point, further investigations of the process of detachment have to be carried out. In the following, we remain in the framework of the "Johnson paradigm" and use the detachment condition (20) and equations based on it. We will describe two types of controlling the external loading conditions: "fixed-grips" and "fixed-load". The loading in the horizontal direction can also correspond either the "fixed-load" or ""fixed-grips" conditions. However, in this paper, we consider only the condition of fixed grips in horizontal direction.

### 2.2.1 Adhesion force under "fixed-load" condition

Under the fixed-load conditions, the instability occurs when the maximum negative normal force is achieved. Thus, the condition of instability can be written as $\mathrm{d}F_z/\mathrm{d}a = 0$. Differentiating Eq. (23) with respect to $a$ and using Eq. (22) we come to the condition

$$\frac{2a_{c,fl}^2}{R}\sqrt{\frac{2\pi\Delta\gamma a_{c,fl}}{E^*} - \frac{G^*}{E^*}u_x^{(0)2}} - \frac{3\pi\Delta\gamma a_{c,fl}}{E^*} + \frac{G^*}{E^*}u_x^{(0)2} = 0 \qquad (28)$$

or, in dimensionless variables (16)

$$\tilde{a}_{c,fl}^2\sqrt{\tilde{a}_{c,fl} - \frac{G^*}{16E^*}\tilde{u}_x^{(0)2}} - \tilde{a}_{c,fl} + \frac{G^*}{24E^*}\tilde{u}_x^{(0)2} = 0. \qquad (29)$$

This equation determines the dependence of the critical radius $\tilde{a}_{c,fl}$ on the tangential displacement $\tilde{u}_x^{(0)}$. The dependence of the adhesion force on the tangential force can be determined by substituting $\tilde{a}_{c,fl}$ into Eqs. (25), (26) and (27). This dependence is shown in Fig. 3a (lower solid line). For $\tilde{F}_x = 0$ (or $\tilde{u}_x^{(0)} = 0$), Eqs. (25), (26) and (29) provide the critical force $\tilde{F}_z(0) = -1$.

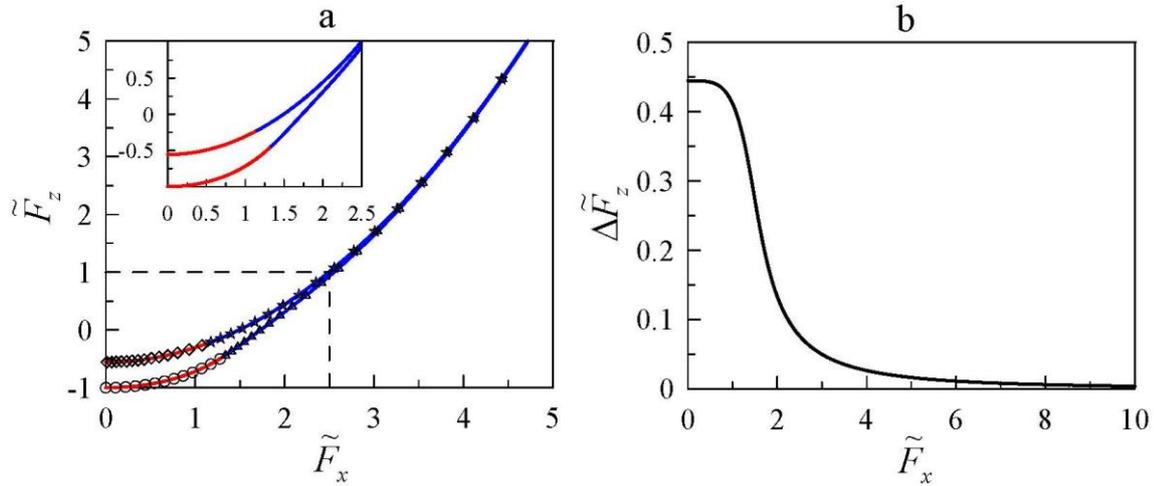

Fig. 3: a) the dependence of normalized critical normal force $\tilde{F}_z$ on normalized critical tangential force $\tilde{F}_x$, for the case $E^* = G^*$. Analytical results are shown by solid lines and results of numerical simulation by open circles, diamonds, stars and triangles. Upper line (diamonds and stars) correspond to the "fixed-grips" conditions in both directions. Lower line (open circles and triangles) corresponds to the "fixed-load" conditions in vertical direction and "fixed-grips" conditions in tangential direction. Diamonds and open circles (red lines) correspond to detaching at a negative indentation depth $d$. Stars and triangles correspond to detaching at a positive indentation depth; b) The difference between normal forces shown in Fig. 3a as function of normalized tangential force $\tilde{F}_x$.

### 2.2.2 Adhesion under "fixed-grips" condition

For the pure normal contact, this mode is shown in Fig. 2a; the detachment occurs at $F_z/F_0 = \tilde{F}_z = -0.5$. The condition of instability in this case formally given by the condition $\mathrm{d}(d)/\mathrm{d}a = 0$. Thus in the general case of superimposed normal and tangential loading, differentiating (25) gives

$$a_{c,fg}^3 - \frac{G^* u_x^{(0)2}}{2\pi\Delta\gamma} a_{c,fg}^2 - \frac{2\pi\Delta\gamma R^2}{16E^*} = 0, \tag{30}$$

or in dimensionless variables (16)

$$\tilde{a}_{c,fg}^3 - \frac{G^*}{16E^*}\left(\tilde{u}_x^{(0)}\tilde{a}_{c,fg}\right)^2 - \frac{1}{9} = 0. \tag{31}$$

This equation determines the dependence of the critical radius $\tilde{a}_{c,fg}$ on the tangential displacement $\tilde{u}_x^{(0)}$. For critical forces at the moment of detachment we obtain

$$\tilde{F}_x^2 = \frac{4G^*}{E^*}\left(\tilde{a}^3 - \frac{1}{9}\right), \tag{32}$$

$$\tilde{F}_z = \tilde{a}^3 - \frac{2}{3}. \tag{33}$$

From the two last equations, it follows

$$\tilde{F}_z = \frac{E^*}{4G^*}\tilde{F}_x^2 - \frac{5}{9}. \tag{34}$$

For the case $E^* = G^*$, the relationship between $\tilde{F}_z$ and $\tilde{F}_x$ under "fixed-grips" conditions simplifies to

$$\tilde{F}_z = \frac{1}{4}\tilde{F}_x^2 - \frac{5}{9}. \tag{35}$$

This dependence is shown in Fig. 3a (upper solid line). Additionally in Fig. 3b the value of difference between normal forces shown in Fig. 3a $\Delta\tilde{F}_z$ are shown as a function of normalized tangential force $\tilde{F}_x$. At zero value of tangential force $\tilde{F}_x = 0$, this difference is equal to $\Delta\tilde{F}_z(0) = \tilde{F}_z^{fg}(0) - \tilde{F}_z^{fl}(0) = -5/9 - (-1) = 4/9$. With increasing $\tilde{F}_x$ the value $\Delta\tilde{F}_z$ monotonically decreases. Note that the lower dependence in Fig. 3a (under "fixed-load" condition in normal direction and "fixed-grips" condition in tangential direction) is well approximated by equation (35) when value $\Delta\tilde{F}_z$ shown in Fig. 3b is close to zero. At $\tilde{F}_x \gg 1$ both dependencies shown in Fig. 3a coincide and are described by equation (35).

## 3  Numerical procedure and comparison with analytical results

In the following, we reproduce the above results numerically and use the designed numerical procedure to expand them to a more complicated detachment condition. Let us first describe briefly the numerical procedure for the case of normal adhesive contact. In the first step, the modified parabolic profile $g(x)$ (6) is indented by $d$ into the elastic foundation shown in Fig. 1b. After this the position of the critical boundary springs (and thus the contact radius) is calculated using the condition (12). The indentation depth and the contact radius given, the normal and tangential forces can be obtained by summing the forces of all springs in contact:

$$F_z = E^* \Delta x \sum_{cont} u_z(x_i),  \tag{36}$$

$$F_x = G^* \Delta x \sum_{cont} u_x(x_i),  \tag{37}$$

where $u_z(x_i)$ and $u_x(x_i)$ are the normal and tangential displacements of individual springs with coordinates $x_i$. In the pure normal contact $u_x(x_i) = 0$, and $F_x = 0$.

When analyzing adhesive contact under "fixed grips" condition, we displace the rigid indenter in the vertical direction step by step with a chosen discretization $\Delta z$. The new configuration of contact after each step is calculated using the rule (12) for the springs at the boundary of the contact. If it happens that there are more than two springs which are detached, we return to the previous step and proceed further with the discretization step $\Delta z / 2$ and decrease it further until only one spring is lost. This procedure is continued up to the point of the instability.

For "fixed-load" conditions the controlling parameter is the normal force $F_{up}$. In each step $F_{up}$ is increased by the increment $\Delta F$, and new equilibrium configuration of the contact is found. Similarly to the procedure for the fixed grips, the increment of the force is decreased if more than two springs are detached in one step. The results of numerical calculations at "fixed-grips" and "fixed-load" conditions at normal motion together with analytical results are presented in Fig. 2. Numerical results coincide with the analytical ones which we can consider as validation of the numerical procedure used.

The procedure in the case of combined normal and tangential loading is basically the same as described above. The essential difference is only the use of a modified detachment condition

$$\Delta l = \sqrt{u_x^2 + u_z^2} = \Delta l_{\max}.  \tag{38}$$

In Fig. 3a the results of numerical calculations are shown along with analytical results presented in the previous Section.

## 4 Conclusion

We performed analytical and numerical simulation of the adhesive contact between two elastic bodies with axially-symmetric gap between them under superimposed normal and tangential loading. The study was done under the simplest assumption that the surface energy does not depend on the detachment direction. However, the developed analytical method can be generalized straightforwardly for more complicated adhesion interactions. Under the above assumptions, the application of tangential force leads to a decrease of the normal adhesive force. We considered different combinations of controlled load and controlled displacement in both normal and tangential direction and derived for each case the "master curves" in appropriate dimensionless variables. The developed method is applicable straightforwardly to adhesive contacts of any bodies of revolution.

**Competing interests.** We declare to have no competing interests.

**Authors' contributions.** V.L.P. proposed the idea of article, have done analytical investigation and wrote the text of article; I.A.L. and A.E.F. carried out numerical analysis of equations and did numerical modeling. All authors gave final approval for publication.

**Funding.** This work was supported in parts by the German Academic Exchange Service (DAAD), by the Ministry of Education of the Russian Federation and by Tomsk State University Academic D.I. Mendeleev Fund Program. I.A.L. grateful to MESU for financial support under the project 0116U006818.